\documentclass[twocolumn,superscriptaddress,secnumarabic,amssymb,amsmath,showpacs,nobibnotes,aps,prl,nofootinbib]{revtex4}
\usepackage{graphicx}
\usepackage{epsf}
\usepackage{bm}
\usepackage{amsmath}
\usepackage{amsfonts}
\usepackage{amssymb}
\usepackage{epstopdf}
\usepackage{natbib}
\usepackage{rotating}
\usepackage{pdflscape}
\usepackage{adjustbox}
\usepackage{color}
\usepackage{dcolumn}
\usepackage{ulem}
\usepackage{times}
\usepackage{appendix}
\usepackage{enumitem}
\usepackage{booktabs}
\usepackage{multirow}
\usepackage{orcidlink}

\begin{document}

\title{Dark energy as generalised superfluid excitations}

\author{El\'ias Castellanos\orcidlink{0000-0002-7615-5004}}
\email{ecastellanos@mctp.mx}
\affiliation{Mesoamerican Centre for Theoretical Physics, \\Universidad Aut\'onoma de Chiapas, Carretera Zapata Km. 4, Real del Bosque (Ter\'an), Tuxtla Guti\'errez 29040, Chiapas, M\'exico.}

\author{Celia Escamilla-Rivera\orcidlink{0000-0002-8929-250X}}
\email{celia.escamilla@nucleares.unam.mx}
\affiliation{Instituto de Ciencias Nucleares, Universidad Nacional Aut\'onoma de M\'exico, Circuito Exterior C.U., A.P. 70-543, M\'exico D.F. 04510, M\'exico.}


\begin{abstract}
In this paper we present a generic form of the cosmological Equation of State derived from the formalism of 
Statistical Mechanics, assuming that a cosmic acceleration scenario can be interpreted as a system 
of quasi--particles. By considering a generalised superfluidity approach, in which the 
energy spectrum can be modulated via a parameter $s$, a negative equation of state 
arises as excitations associated with an \textit{exotic} superfluid system.
We show that for $s\sim 0.293$, $w\simeq-1$, which can be related to the standard equation of state for the $\Lambda$CDM model.
\end{abstract}

\maketitle

\section{From Statistical Mechanics to standard FLRW cosmology}
The observed dynamics of our universe is often described in terms of the FLRW metric. Even though, the adoption of this
convenient geometry in Einstein's equations does not give information about the \textit{specific} form for the
Equation of State (EoS) of the effective cosmological fluid, $P=w \rho_{E}$. Here, 
the total energy density in the universe is denoted by $\rho_{E}$ and the pressure by $P$.
The standard concordance model, $\Lambda$CDM, fits in this scenario where the fluid can be identified 
into
matter (including baryonic and cold dark matter) $\rho_m$, radiation $\rho_r$, and a not so well understood dark energy $\rho_{\Lambda}$.
Towards this direction, many
attempts have been done starting from the simple relation above described by a barotropic EoS (notice
that here we call $w$ as the EoS parameter) to some complex ones with an explicit relation between $P$ and $\rho_{E}$. All of them to 
achieve the current cosmic acceleration. As it is standard, to obtain this behaviour we require an energy density 
with significant negative pressure at late times. This means that the evolution ratio between the pressure and energy density is negative, i.e.
$w(z)= P/\rho_{E} <0$. All reasonable fitting dark energy models available in the literature are in agreement at this point \cite{Riess:1998cb,Garnavich:1998th,Tonry:2003zg,Padmanabhan:2002ji,Peebles:2002gy,Caldwell:1997ii}.

The evolution Friedmann equation for a spatially flat universe
\begin{eqnarray} \label{eq:hubble_eq}
 \left(\frac{H(z)}{H_0}\right)^2 
 &\propto& 
 \left[\Omega_{m} (1+z)^3 +\Omega_{\Lambda} f(z)\right], \quad~
 \end{eqnarray}
where $H_0$ is the Hubble parameter and the index $m$ denotes the matter components related to radiation, barionic and cold dark matter. From this equation it is possible to compute the EoS, $w(z)$. This can be done when
the current value for the dark energy density is written as $\rho_{0(\Lambda)}=\rho_{\Lambda} (z) f^{-1}(z)$, with $f(z)= \text{exp}\left[3\int^{z}_{0}{\frac{1+w(\tilde{z})}{1+\tilde{z}}}d\tilde{z}\right]$. By modelling $w(z)$, we can give directly an entire evolution description
of Eq.~(\ref{eq:hubble_eq}), as e.g., in the case of
quintessence models $w=$constant, the solution for $f(z)$ is $f(z)=(1+z)^{3(1+w)}$. 

If we consider the case of the Cosmological 
Constant $\Lambda$ ($w=-1$) we obtain $f =1$. Other cases explore a dark energy density $\rho_{\Lambda}$ with 
varying and non-varying $w(z)$ \cite{Shafieloo:2012rs,Chevallier:2000qy,Escamilla-Rivera:2016qwv,Clifton:2006jh}, just to cite a few.

Notice that for a given functional form of $f(z)$, the contribution
of the dark energy density to $H(z)$, goes to more 
negative values of $w(z)$. This is an impact in the evolution
of dark energy on the dynamical age of the universe.
As we mentioned, to get a dark energy model with late-time negative pressure
we can think in two frameworks: on one hand, a quintessence model which shows a wide 
application in tracker the slow roll condition of scalar fields\footnote{An associated particle can be identified with a boson with zero spin.} and
demands a constant EoS \cite{Nesseris:2004wj}, a references therein. In numerical terms, according to Planck 2018 \cite{Aghanim:2018eyx},
the dark energy EoS parameter for a flat universe is $w=-1.006\pm 0.045$, which is 
consistent with $\Lambda$. 
On the other hand, for kinessence  models, the EoS is a function of $z$ and several dark energy models 
with different parameterisations of $w(z)$ has been discussed in the literature \cite{Sahni:2002fz}.

According to the latter ideas, several types of \textit{fluid} are used for cosmological viable scenarios, as:
\begin{itemize}
\item Non--relativistic matter: An ordinary non-relativistic matter EoS  as $w=0$ (i.e cold dust), correspond to a diluted fluid with $\rho \propto a^{-3}=V_{3}^{-1}$, where $V_{3}$ is the volume in 3 space--like dimensions.
\item Ultra--relativistic matter: An ultra-relativistic matter EoS given by $w=1/3$ (e.g. radiation, or matter in the very early universe), correspond to a fluid as $\rho \propto a^{-4}$. In an expanding universe, the energy density decreases faster than the volume expansion, because radiation has a momentum and, by the de Broglie hypothesis a wavelength is redshifted.
\item Cosmic accelerated inflation: This case can be characterised by a dark energy EoS. In the simplest scenario, this correspond to the EoS of $\Lambda$ with $w=-1$. The solution of the conservation equation for the scale factor is not valid and $a\propto e^{Ht}$. In general, the expansion of the universe is accelerating for any EoS $w< -1/3$. 
\item Hypothetical fluids: In this case, a hypothetical phantom energy is considered to be with a EoS $w<-1$, and we will have a Big Rip. Using current data, it is impossible to distinguish between phantom $w<-1$ and non-phantom $w\geq -1$.
\end{itemize}
We should mention that the first two cases can be derived from first principles, since that can be deduced from quantum and statistical mechanics. The dark energy and phantom--like fluids cannot be derived as the latter. From the standard model conception, it seems that in our universe exist only two types of particles: bosons and fermions \footnote{Dark matter and dark energy could be in a different scheme. However, we assume here that even in this case there are only bosons and fermions in the universe.}. The nature of this difference relies basically on some intrinsic property, the so--called spin. As a consequence, the statistics, i.e., how these particles occupy a single quantum energy state, depends on their quantum nature. Furthermore, it is well known that the system under consideration displays a significant departure from their classical behaviour at certain critical temperature, depending on whether particles are the constituents of the system (bosons or fermions). In other words, the quantum behaviour is relevant, and under these circumstances and the quantum properties of the corresponding particles must be taken into account. Bosons obey Bose--Einstein statistics and fermions obey Fermi--Dirac statistics. Summarising, the corresponding properties of both systems are quite different when quantum properties are taken into account, the bosons do not fulfill the Pauli's exclusion principle, while fermions do. 

Thus, according to the concepts described above, it can be demonstrated that the occupation number in the Grand Canonical Ensemble can be expressed as
\footnote{Defined also as the partition function. In this work we assume the equivalence among different ensembles, i.e., our system lies in the thermodynamic limit. In consequence, the choice of another ensemble is irrelevant for the approach followed in the present work since it predicts the same physics.}
\begin{equation}
\label{eq:on}
n(\epsilon_{p})_{\pm}=\frac{1}{\exp{\beta(\epsilon_{p}-\mu)}\pm1},
\end{equation}
where $(+)$ stands for fermions and $(-)$ for bosons. As usual, $\beta=1/\kappa T$, being $\kappa$ the Boltzmann's constant and $T$ the temperature. Additionally, $\epsilon_{p}$ is the corresponding single--particle energy spectrum and $\mu$ is the chemical potential. Thus, we have to deduce, through the dynamics, what is the specific functional form of the energy spectrum for a single particle to analyse the occupation number Eq.\,(\ref{eq:on}), which contains relevant thermodynamical information of the system. 
At this point, it is important to mention that the corresponding EoS contains, in fact, the aforementioned relevant thermodynamical information, if one knows the dynamical properties of the system, i.e., if these properties can be deduced from the experiment or a microscopic model of matter. In other words, the analytical form of the EoS depends on the substance under consideration, and its deduction is dependent on the experiment and/or a microscopic  model of matter \cite{Gar_colin}. 

Let us calculate the EoS for some thermodynamical systems, by using all the information given above. For this goal, let us appeal to Ref.\,\cite{pathria}
 to exemplify the method. We assume that the semiclassical single--particle energy spectrum for a given system can be expressed as $\epsilon=p^{s}/\alpha$, where $p$ is the corresponding momentum, $s$ is some positive real number and $\alpha$ is a constant with an adequate dimensions. The system under consideration is ideal, i.e., interactions among its constituents are neglected. To generalise, our system can be  situated in $n$ space--like dimensions. Under these circumstances it is straightforward to show that the EoS is given by \cite{pathria}
\begin{equation}
\label{eq:es}
P=\frac{s}{n}\left(\frac{U}{V_{n}}\right),
\end{equation}
where, $P$ is the pressure, $U$ is the internal energy, and $V_{n}$ is the hyper volume under consideration. Thus, we can define the energy density for any space--like dimension as $\rho_{\epsilon}=(U/V_{n})$.

By using the general relation Eq.\,(\ref{eq:es}), we are able to recover the standard results in 3 space--like dimensions $(V_{3})$: 

\begin{itemize}
\item Non--relativistic system:  EoS  of the form  $P=\frac{2}{3}(U/V_3)=\frac{2}{3}\rho_{\epsilon}$ corresponding to a single--particle energy spectrum $\epsilon= \frac{p^{2}}{2m}$.

\item Ultra--relativistic system: the EoS is this case is of the form  $P=\frac{1}{3}(U/V_3)=\frac{1}{3}\rho_{\epsilon}$ which corresponds to the single--particle energy spectrum $\epsilon= c\,p$, with $c$ the speed of light.
\end{itemize}
Let us remark that the properties of the system, and in particular the EoS, strongly depends on the functional form of the single--particle energy spectrum.
Due to the functional form of the single particle energy spectrum $\epsilon \sim p^{s}$, we 
can obtain
an infinite set of EoS in 3 space--like dimensions, depending on the value of the parameter $s$. However, the last assertion can lead to \textit{exotic} single--particle energy spectra which can be related, in principle, to the kinetic energy of each particle. 

Finally, notice that in the case of radiation the EoS is given by $P=\frac{1}{3}(U/V_3)=\frac{1}{3}\rho_{\epsilon}$. Nevertheless, in this case, we have to make some additional assumptions because for these systems, the total number of particles is not conserved, and consequently, the corresponding chemical potential must be set equal to zero \cite{pathria}. Moreover, let us remark that the EoS for radiation (together with the corresponding EoS for dust $P=0$) can be obtained also from first principles without deeper conceptual complications.

Finally, we focus on the following scenarios for different values of the parameter $s$. For instance, notice that in 3 space--like dimensions when $s=3$, the relation between the pressure and the energy density is such that $P=(U/V_3)=\rho_{\epsilon}$, which is a consequence of a semiclassical single--particle energy spectrum of the form $\epsilon= p^{3}/\alpha$. 
Notice that an EoS of the form
$P=-(U/V_3)=-\rho_{\epsilon}$ (in the case of dark energy) implies a single--particle energy spectrum $\epsilon=p^{-3}/\alpha$ by using Eq.~(\ref{eq:es}). The last assumption has no physical meaning, since a dispersion relation $ \sim p^{-3}$, leads to a negative number of micro--states due to the above analysis is only valid for real positive values of $s$ and it is unable to reproduce the standard EoS for dark energy.

\section{Superfluid excitations as dark energy EoS}
To put in context the system under consideration, let us first analyse some thermodynamical conditions. For this goal, we will describe the nature of the system associated with our model. The system under consideration is a picture of a condensate of some generic bosons (or scalars) and its corresponding elementary excitations.
In this aim we calculate the condensation temperature in order to analyse the possibilities to get a condensed state. We start with the semiclassical single--particle energy spectrum $\epsilon=p^{s}/\alpha$, in three space--like dimensions for a real positive number $s$.

In this scenario the condensation temperature $T_{c}$ is given by
\begin{equation}
\label{eq:TC}
T_{c}=\frac{1}{\kappa}\left[ \left( \frac{s\,h^{3}  \Gamma(3/s)}{4\pi \alpha^{3/s}\zeta(3/s)}\right) n\right]^{s/3},
\end{equation}
where, $n=N/V$ is the density of particles, $\kappa$ is the Boltzmann's constant, $h$ is the Planck's constant, $\Gamma(3/s)$ is the Gamma Function and $\zeta(3/s)$ is the Riemann Zeta Function. Let us remark that for specific values of the parameter $s$ in the range of $0<s<3$ the condensation temperature is well defined in the thermodynamic limit.
In other words, the temperature of the system in which we are able to describe the associated excitations corresponds at least to temperatures $T<T_{c}$ which is well defined for values of the parameter $s$ between $0<s<3$. This last assertion restringes the onset of condensation and consequently the value of the parameter $s$, at least in the thermodynamic limit by fundamental physics.

In this line of thought, we propose that a generalised Landau's roton spectrum for dark energy in 3 space--like dimensions that can be written as follows
%
\begin{equation}
\label{eq:Nes}
\epsilon_{p}=\Delta + \frac{(p-p_{0})^s}{\alpha},
\end{equation}
where  $s$ is a real positive number. This 
energy spectrum is a generalisation of the spectrum associated with  superfluidity behaviour of liquid Helium II when $s=2$, or the so--called Landau's roton spectrum \cite{pathria,pethick}. In Eq.\,(\ref{eq:Nes}), $\Delta=\epsilon_{p=p_{0}}$ is some gap energy. 
The term $p_{0}$, can be interpreted as the momentum near to the extremal of the energy spectrum Eq.~(\ref{eq:Nes}), i.e., has the behaviour of a quasi--particle. Excitations with momenta close to $p_0$ are referred to as rotons in the usual non--relativistic case, i.e., for $s=2$, when the chemical potential $\mu$ is set to be zero.
We must mention here that a typical superfluid behaviour of the system at low momenta has a linear dispersion relation $\epsilon_{p}=c\,p$ with $c$ the speed of sound when $s=2$ see for instance \cite{pathria,pethick}. This linear dispersion relation at low momenta is caused by the presence of interactions within the system that create elementary excitations related somehow to the so--called Bogoliubov excitations \cite{maeda}. Thus in the case $s=2$, the excitation spectrum at the low--energy limit ($p \rightarrow 0$) is identical to that of a phonon, where the sound velocity $c$ is given by $c=\frac{\hbar}{m}\sqrt{4\pi a\,n}$, being $a$ the s--wave scattering length and $n$ the corresponding density. Excitations with momenta in the linear region are called phonons; those with momenta close to the minimum are called rotons. Excitations with momenta near the maximum are called maxons. However, when $0<s<3$ with $s\neq2$ at very low momenta this situation should change according to the specific value of the parameter $s$, see Appendix A.

We compute the relation between the pressure $P$ and the internal energy $U$ associated with 
Eq.\,(\ref{eq:Nes}) near to $p_{0}$, in other words, the corresponding EoS. Consequently, we obtain \footnote{From this point forward,  the volume $V$ is assumed to be in 3 space--like dimensions.} 
\begin{equation}
\label{eq:P1}
\frac{PV}{\kappa T}=\frac{4\pi V}{h^{3}}\int_{0}^{\infty}e^{-\Bigl[\Delta+\frac{(p-p_{0})^{s}}{\alpha}\Bigr]/\kappa T}p^{2}dp\simeq\overline{N},
\end{equation}
where $\overline{N}$ is the \textit{equilibrium} number of quasi--particles that we call for our purposes \textit{generalised rotons}. 

To solve the integral Eq.\,(\ref{eq:P1}) we take, as was mentioned above $s>0$, together with the assumption that for low temperatures, i.e., for temperatures of interest, the minimal value of the occupation number $n(\epsilon_{p})$ is at most $\exp{(-\Delta/\kappa T)}$ which is larger than unity. Thus, we can express the occupation number Eq.\,(\ref{eq:on}) for bosons as $n(\epsilon_{p})\approx \exp{(-\beta \epsilon_p)}$.\\ With the change of variables  $p=p_0 + (\alpha \kappa T)^{1/s} x$, we get from Eq.\,(\ref{eq:P1})
\begin{eqnarray}
\frac{PV}{\kappa T}&=&\frac{4\pi V}{h^{3}}p_{0}^{2}(\alpha\kappa T)^{1/s}e^{-\Delta/\kappa T}
\nonumber \\ &&
\times \int_{-\infty}^{\infty} e^{-s}\left(1+\frac{(\alpha\kappa T)^{1/s}}{p_{0}}x+\frac{(\alpha\kappa T)^{2/s}}{p_{0}^{2}}x^{2}\right), \quad \quad
\end{eqnarray}
without losing generality we are able to obtain
\begin{equation}
\label{eq:PV}
\frac{PV}{\kappa T}=\frac{4\pi V}{h^{3}s}\left(\alpha \kappa T\right)^{1/s}p_{0}^{2}e^{-\Delta/\kappa T}f(s)
\end{equation}
where
\begin{eqnarray}
f(s)&=& 3- \Gamma\left(\frac{1}{s}\right) -\frac{(\alpha \kappa T)^{1/s}}{p_0} \Gamma\left(\frac{2}{s}\right) 
\nonumber \\&&
-\frac{(\alpha \kappa T)^{2/s}}{p_0} \Gamma\left(\frac{3}{s}\right), ~
\end{eqnarray}
and $\Gamma(\mu)$ is the Gamma function. 
To achieve the corresponding internal energy $U$, we employ the useful thermodynamic functions $A= -PV$ (Helmholtz free energy) and $S=-\left(\frac{\partial A}{\partial T}\right)_{V}$ (entropy), with $U=A+TS$.

At this point it is important to mention that the phonon contributions to the various thermodynamic properties of the system become rather unimportant near to $p_{0}$; then, the rotons are the only excitations that need to be considered. Thus, according to our model the particles that form the condensed phase does not contribute to the cosmological energy density and pressure. After straightforward calculations, we obtain  for the energy density $\rho_{E}$
\begin{eqnarray}
\frac{U}{V}&\equiv&\rho_{E}=\frac{4\pi }{h^{3} s^{2}} e^{-(\Delta/\kappa T)}p_{0}^{2} (\alpha \kappa T)^{1/s} 
\Big\{ 3(\Delta s +\kappa T)
\nonumber \\ && 
- (\alpha \kappa T)^{1/s} 
\Big[2 (\Delta s +s \kappa T) \Gamma\left(\frac{1}{s}\right) - (\alpha \kappa T)^{1/s} 
\nonumber \\ &\times&
\Big(- (\Delta s +3s \kappa T)\Gamma\left(\frac{2}{3}\right) 
\nonumber \\ &&
+ (\alpha \kappa T)^{1/s} (\Delta s +4\kappa T)\Gamma\left(\frac{3}{s}\right)  \Big) \Big]  \Big\}.
\label{eq:IE}
\end{eqnarray}
Multiplying and dividing by $s (\kappa T f(s))$ we get
\begin{equation}
\label{eq:UP}
P=s \rho_{E}\left(\frac{f(s)}{g(s)}\right),
\end{equation}
where  $g(s)$ is given by
\begin{eqnarray}
g(s)&=& \frac{1}{\kappa T} \Big\{ 3(\Delta s +\kappa T)- (\alpha \kappa T)^{1/s} \Big[2 (\Delta s +s \kappa T) 
\nonumber \\ &\times&
\Gamma\left(\frac{1}{s}\right) - (\alpha \kappa T)^{1/s} \Big(- (\Delta s +3s \kappa T)\Gamma\left(\frac{2}{3}\right) 
\nonumber \\ &&
+ (\alpha \kappa T)^{1/s} (\Delta s +4\kappa T)\Gamma\left(\frac{3}{s}\right)  \Big) \Big]  \Big\}.
\end{eqnarray}
It must be mentioned that at $T=0$, the above results tend to zero and clearly, there are not \textit{generalised rotons} in the system.
Notice also that for even values of the parameter $s$ the term $s\left(\frac{f(s)}{g(s)}\right)$ is always positive. The interesting scenarios for the EoS Eq.\,(\ref{eq:UP}) are given for odd values of the parameter $s$, together with values between $0<s<1$, in which the right hand term of Eq.\,(\ref{eq:UP}) becomes negative. This term seems to be relevant in the search for negative contributions on the energy density of the EoS Eq.\,(\ref{eq:UP}) near to $p_{0}$, that we interpreted as dark energy.
\begin{figure*}
\centering
\includegraphics[width=0.48\textwidth,origin=c,angle=0]{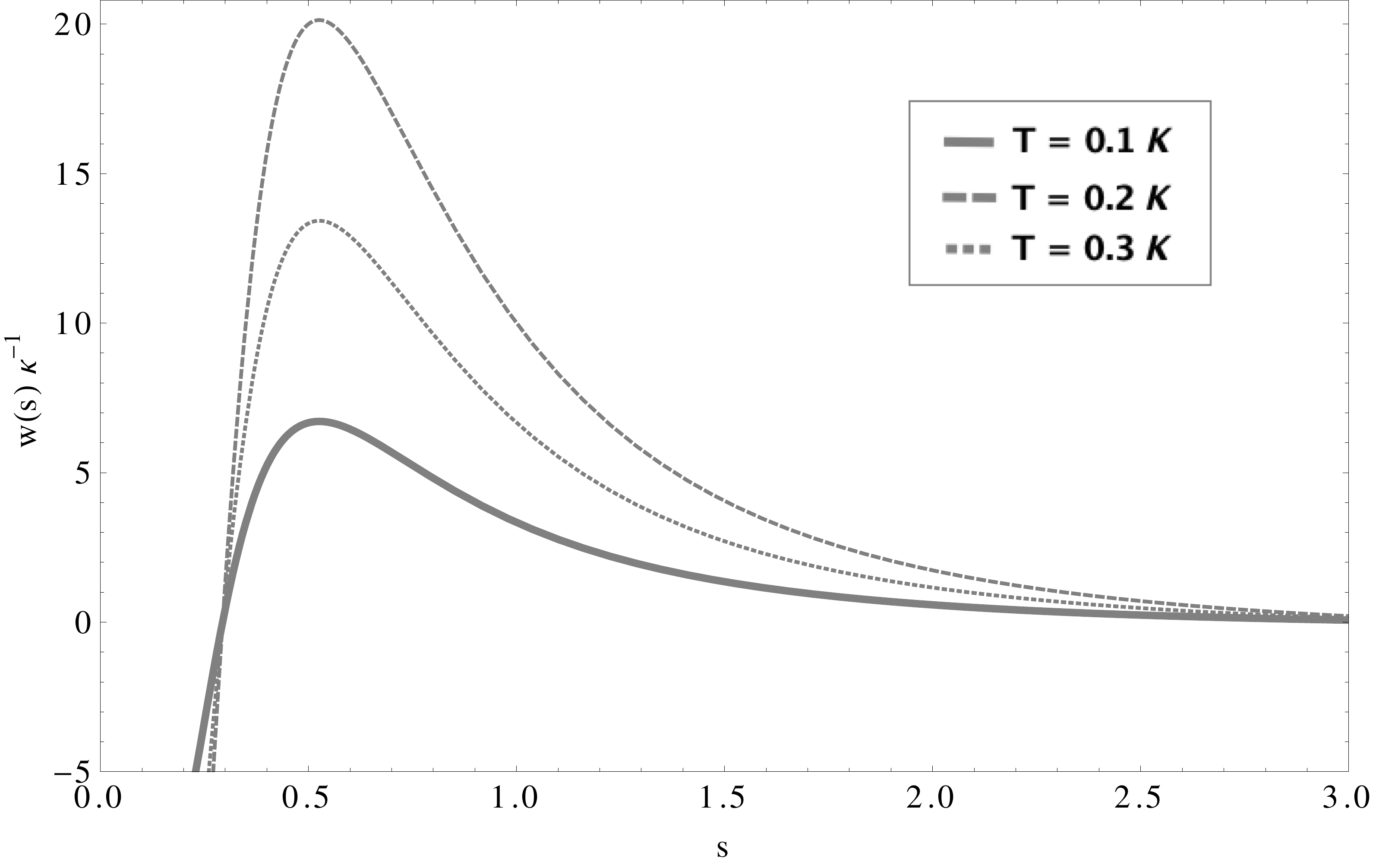}
\includegraphics[width=0.50\textwidth,origin=c,angle=0]{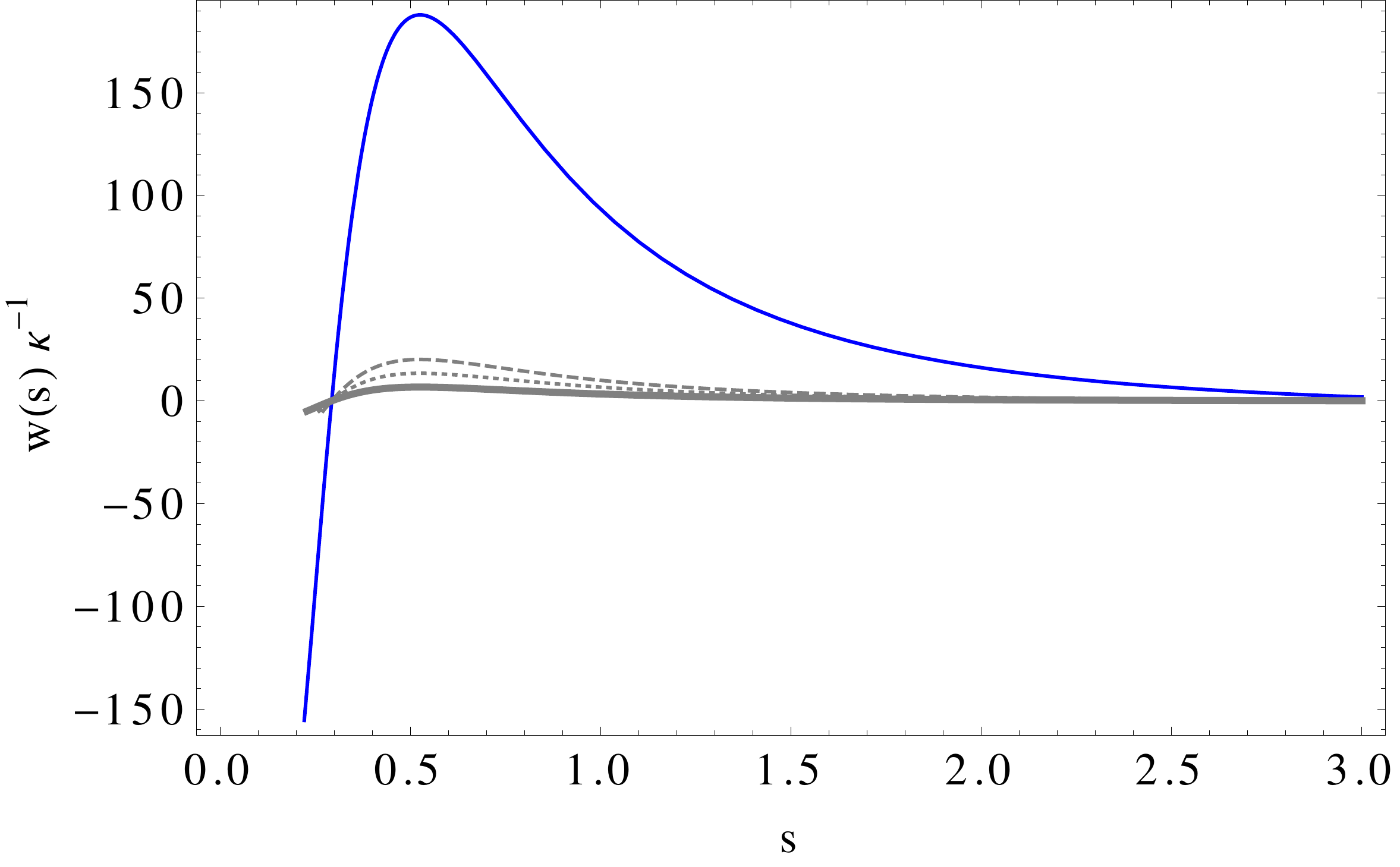}
\includegraphics[width=0.47\textwidth,origin=c,angle=0]{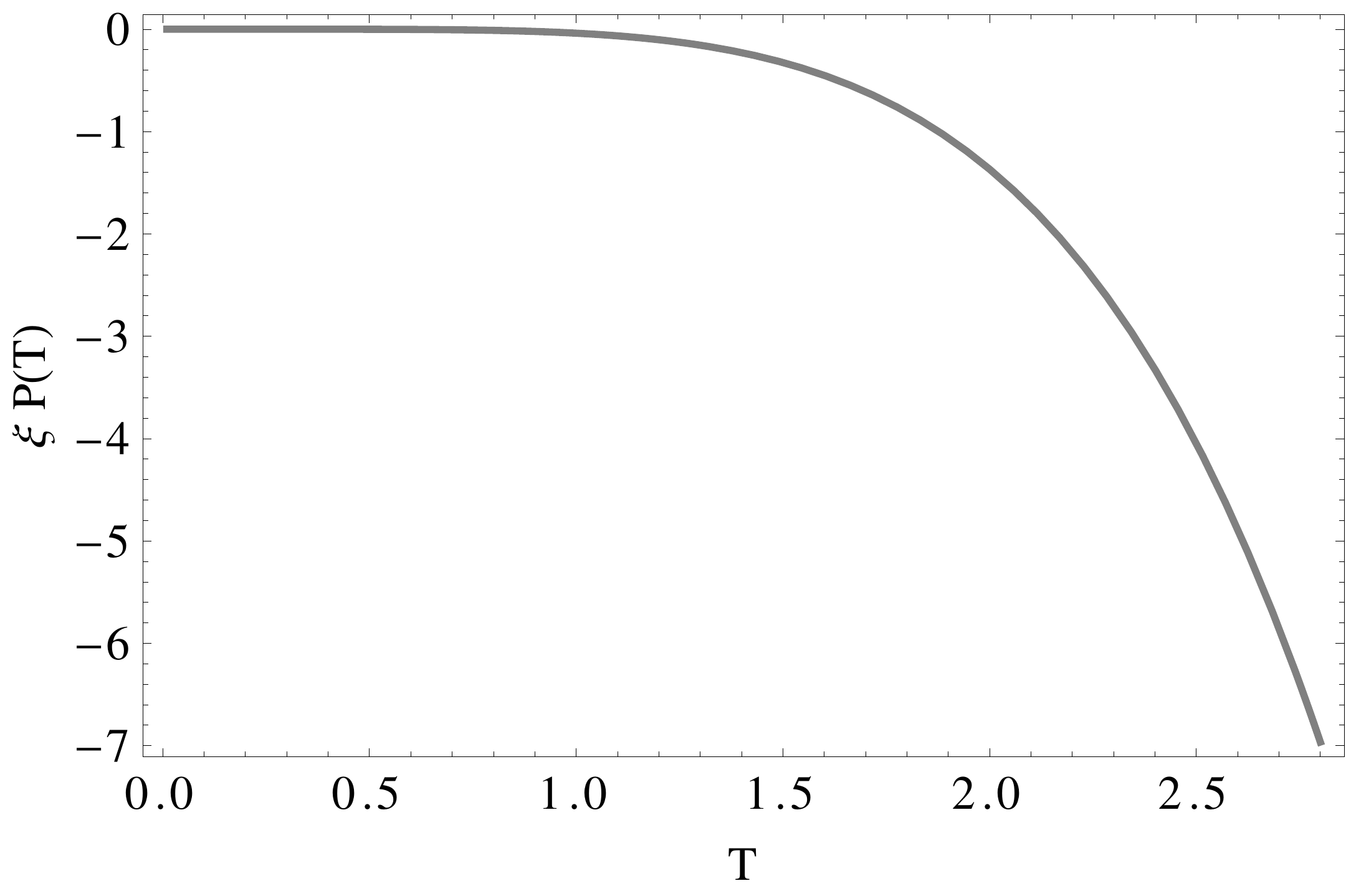}
\caption{\textit{Top left:} Plot of $w(s)$ in terms of $f(s)$ and $g(s)$ at different low temperatures in Kelvins (K) indicated inside the box plot. The EoS units considered here are normalised by $\kappa$. The values taking into account were: $\alpha =1.1$, $\Delta = 0.1$, $\kappa=1.3\times 10^{-23}$ and $p_0 =1.0$.
It is interesting to notice that all the graphs intersect when $s\sim 0.293$, which corresponds to $w(s)\simeq -1$.
\textit{Top right:} Plot of $w(s)$ at the order of the CMB temperature (blue curve) $T\approx 2.8$ K. Notice that the intersection value of $s$ remains unchanged.
\textit{Bottom:} The pressure from Eq.~(\ref{eq:PV}). We consider first order terms in $f(s)$ and the obtained value $s=0.293$. The pressure was normalised in units of $\xi = 10^{-13} e^{\Delta /\kappa} h^{3} \kappa^{-1}.$
Clearly, while the temperature tends to zero the pressure vanishes, which corresponds to an EoS as dust ($w\rightarrow 0$).
}
\label{fig:EoSTermo}
\end{figure*}

We define our EoS Eq.\,(\ref{eq:UP}) in terms of $ w$ and $s$ as 
\begin{equation}
\label{eq:EOSF}
w(s)\equiv\frac{P}{\rho_{E}}= s\left(\frac{f(s)}{g(s)}\right),
\end{equation}
with units of $[T]=$\,Kelvin, $[\Delta] =$\,Joules, $\alpha=$\, seconds/meter, $[p_0]=$\,kilograms\,meters/seconds and, $\kappa = 1.3 \times 10^{-23}\text{Joules/Kelvin}$. \\
In Figure\,(\ref{fig:EoSTermo}) we plot Eq.\,(\ref{eq:EOSF}) in function of the parameter $s$ at low temperatures. 
We have three relevant regions which depict the behaviour of the system depending on the value of $s$. Specifically, when $s$ is large enough we have that $w(s) \rightarrow 0$, i.e., the system behaves as dust. We obtain that for values $s \gtrsim 0.293$ the system behaves as a \textit{normal fluid}, in the sense that the energy density is positive. The interesting scenario seems to be placed for values $s \lesssim 0.293$ in which the energy density becomes negative. For $s\lesssim0.293$ the system behaves as phantom fluid with $w(s)<-1$. For $s \sim 0.293$, the value of $w(s)$ approaches to $-1$ in a range of low temperatures according to our approximation. Using the facts described above, we can rewrite Eq.~(\ref{eq:Nes}) with $s \sim 0.293$ as $\epsilon_{p} \sim (p-p_{0})^{0.293}$. Therefore, in our approach the excitations (or quasi--particles) that we call \textit{generalised rotons}, close enough to $p_{0}$ of the corresponding energy spectrum Eq.~(\ref{eq:Nes}), are able to mimic dark energy when $s \sim 0.293$.\\
At this point we must mention that an optimal path to fix the free parameters of our model is to evaluate Eq.\,(\ref{eq:Nes}) by considering the initial value $p_0 \geq 0$, in order to obtain initially the case $\epsilon \propto p^s/\alpha$. Once we consider our ansatz Eq.\,(\ref{eq:Nes}) we select a maximum value around the CMB temperature ($T=2.8$K) (see also Figure \ref{fig:EoSTermo}; which in fact must be smaller that the corresponding critical temperature $T_{c}$ \,Eq.\,(\ref{eq:TC})), where the expression inside the parenthesis in Eq.\,(\ref{eq:Nes}) is valid and the values for $\alpha$ does not change significantly. It seems to be that according to temperatures of order of the CMB, Eq.\,(\ref{eq:TC}) predicts around $10^{143}$ \textit{condensed particles} in the universe\footnote{We consider a current expansion rate that indicates a critical density of the universe about $9\times 10^{-27} \text{kg} \text{m}^{-3}$. However, this density is the total density of both matter and energy.}.
For this case, the only possible initial value is $\alpha=1.1$ and then $p_0 \lesssim 10^{-12}$ by using the precedent values. We must mention here that we are not able to obtain the corresponding values of the free parameters from first principles or from fundamental physics \footnote{However, we can fix an interval for the parameter $s$ obtained in Eq.\,(\ref{eq:TC}).}. At this level the justification for the election of these specific values for the free parameters is precisely that we are not able to get an EoS of the form $P=-\rho_{E}$ outside the range of vales presented above. Additionally, the relevant quantities for our analysis will be rescaled for an interval between $p:\{0,1]$. Clearly, we must confront the predictions of the model also with observations in order to get some insight to bound the corresponding free parameters. These topics deserve deeper analysis that we will present elsewhere.




\section{Discussion}
In this paper, we present an EoS described by a generic Landau's roton spectrum in 3 space--like dimensions, which can reproduce dark energy characteristics in certain laboratory conditions.
To achieve this condition, we define a free parameter $s$ that modulates the $(p-p_0)$ term that appears in Eq.\,(\ref{eq:Nes}), which value can be used to obtain direct information of the corresponding EoS for the superfluid. We obtain that the value $s\sim 0.293$, reproduces $w(s)\simeq -1$. This result leads to a scenario that mimics a $\Lambda$CDM EoS that can be eventually extended (and perhaps tested) with superfluidity in the laboratory. 
These type of \textit{fractional powers} of $s$, are not an unusual topic in modern physics, e.g., they are related to properties of fractional derivatives and consequently, with \textit{fractional kinetics} \cite{rh,fr,gm,md}. Even more, they can be used in several physical applications, see for instance \cite{gm,md,rhh,ka} and references therein.
For instance, in Ref.\,\cite{rh} it was analysed a fractional Schr\"odinger--type wave equation in 1--dimension for a free particle. In this case, the corresponding energy spectrum is given by 
$\epsilon_{k}=\eta (\hbar k)^{2\gamma}$, where $\gamma$ takes real positive values and $\eta$ is defined as $\eta \equiv \frac{1}{2}(m^{2\gamma-1} c^{2(\gamma-1)})^{-1}$. It is noteworthy to mention that this result is in agreement with our approach when
$s=2\gamma$. \\
According to the results obtained in the Appendix A, the system behaves as $p^{1/6}$, in which we interpreted the term $\sim \alpha^{1/2}\sqrt{\frac{2U_{0}N}{V}}$ as the \textit{speed of sound} for low momenta up to some maximal value $p_{m}$. Additionally, it is seem to be that there is a region between $p_m$ and $p_{0}$ in which we do not have any information about the functional form of $\epsilon_{p}$. In fact, if we assume the usual behaviour, this region must be formed by a combination of phonos and rotons. Thus, it could be interesting to explore the system in this interval of values for the momenta for low temperatures (or even a scenario of dynamical $s$) in oder to extract information of the system and set at least bounds for the free parameters of the model and eventually compare with observations. These issues are out of the scope of this work, however will be explored in future works.
Finally, it is remarkable that our proposal gives insights that can eventually be applied at cosmological scales to understand the nature of dark energy in the universe.

\bigskip
\textbf{Acknowledgments:}
EC acknowledges the receipt of the grant from the Abdus Salam International Centre for Theoretical Physics, Trieste, Italy. CE-R acknowledges the Royal Astronomical Society as FRAS 10147, 
PAPIIT Project IA100220 and would like to acknowledge networking support by the COST Action CA18108.


\bigskip
\bigskip
\appendix{\bf{Appendix A}}\\

Let us analyse the corresponding excitation spectrum at the low--energy limit $p \ll p_{0}$, i.e., the phonon--like behaviour, if there exist,  for the specific value $s\sim 1/3$. Let us start with a N--body Hamiltonian in three dimensions where we has just introducing $p=\hbar k$
\begin{eqnarray}
\hat{H}=\sum_{\vec{k}=0}\Bigg[\frac{\hbar^{s}k^s}{\alpha}
\Bigg]\,\, \hat{a}_{\vec{k}}^{\dagger}\hat{a}_{\vec{k}}+\frac{U_0}{2V}\sum_{\vec{k}=0}\sum_{\vec{p}=0}\sum_{\vec{q}=0}\hat{a}_{\vec{p}}^{\dagger}\hat{a}_{\vec{q}}^{\dagger}
 \hat{a}_{\vec{p}+\vec{k}}\hat{a}_{\vec{q}-\vec{k}} , ~~ \label{Ham1}
\end{eqnarray}
where the $U_{0}=4\pi \hbar^{2} a/m$ is the corresponding interaction potential with $a$ the scattering length. Additionally, $\hat{a}_{\vec{k}}^{\dagger},\,\hat{a}_{\vec{k}}$ are the corresponding creation and annihilation operators that satisfy the usual canonical commutation relations for bosons.
%
We assume that for temperatures $T<T_{c}$ (see Eq.\,(\ref{eq:TC}))
\begin{equation}
N_{0}\approx N, \,\,\,\,\,\,\,\, \sum_{\vec{k} \not=0} N_{\vec{k}} <<N,
\end{equation}
being $N$ the total number of particles, $N_{\vec{k}}$ the number of particles in the excited states, and $N_{0}$ the number of particles in the ground state. Keeping terms up to second order in $\hat{a}_{0}$, the Hamiltonian (\ref{Ham1}) becomes
\begin{eqnarray}
\hat{H} &=&\frac{U_0N^2}{2V}+
\sum_{\vec{k}\not=0}\Bigl[\frac{\hbar^s k^s}{\alpha}+\frac{U_0N}{V}\Bigr]\hat{a}_{\vec{k}}^{\dagger}\hat{a}_{\vec{k}}\nonumber\\&+&2\hat{a^{\dag}_{0}} \hat{a}_{0} \sum_{\vec{k}\not=0}\hat{a}_{\vec{k}}^{\dagger}\hat{a}_{\vec{k}}+\hat{a_{0}^{\dag^{2}}}\sum_{\vec{k}\not=0}\hat{a}_{\vec{k}}\hat{a}_{-\vec{k}}
+ \hat{a_{0}}^{2} \sum_{\vec{k}\not=0 }\hat{a}_{\vec{k}}^{\dagger}\hat{a}_{-\vec{k}}^{\dagger}.\,\,\,\,\,\,\,\,~~
\label{Add2}
\end{eqnarray}
Finally, in the same order of the approximation we assume that $\hat{a}_{0}^{\dag} \hat{a}_{0},\ \hat{a}_{0}^{2},\ 
 \hat{a_{0}^{\dag^{2}}} = N$. Using these facts, the Hamiltonian (\ref{Ham1}) can be re--expressed as follows
\begin{eqnarray}
\hat{H} &= &\frac{U_0N^2}{2V}+\sum_{\vec{k}\not=0}\Bigl[\frac{\hbar^sk^s}{\alpha}+\frac{U_0N}{V}\Bigr]\hat{a}_{\vec{k}}
^{\dagger}\hat{a}_{\vec{k}}\nonumber\\
&+&\sum_{\vec{k}\not=0}\frac{U_0N}{2V}\Bigl[\hat{a}_{\vec{k}}^{\dagger}\hat{a}_{-\vec{k}}^{\dagger}
+ \hat{a}_{\vec{k}}\hat{a}_{-\vec{k}}\Bigr]. \label{Ham2}
\end{eqnarray}
In order to  obtain the ground state energy associated with our system, let us diagonalize the above Hamiltonian by introducing  the so--called Bogoliubov
transformations \cite{pethick,pathria}.

After some algebra, the following diagonalized Hamiltonian can be obtained
\begin{eqnarray}
\label{Ham3}
&&\hat{H}=\frac{U_0N^2}{2V}+ \sum_{\vec{k}\not=0}\sqrt{\epsilon_{k}\Bigl(\epsilon_{k}+\frac{2U_0N}{V}\Bigl)}
\hat{b}_{\vec{k}}^{\dagger}\hat{b}_{\vec{k}}\\
&+&\sum_{\vec{k}\not=0}\Bigg\{-\frac{1}{2}\Bigg[\frac{U_0N}{V} +\epsilon_{k}
-\sqrt{\epsilon_{k}\Bigl(\epsilon_{k}+\frac{2U_0N}{V}\Bigr)}\,\,\Bigg]\Bigg\}, \nonumber
\end{eqnarray}
where $\epsilon_{k}=\frac{\hbar^sk^s}{\alpha}$. The last summation in the Hamiltonian (\ref{Ham3}) diverges as $(U_{0}N/V)^{2}/2\epsilon_{k}$, as can be seen by performing an expansion of the last term in Eq.\,(\ref{Ham3}) for large $k$.

 The pseudo--potential $U_{P}(\bold{r})$ can be expressed as follows
 \begin{equation}
 \label{PP}
 U_{P}(\bold{r})=U_{0}\delta(\bold{r})\frac{\partial}{\partial r}r,
 \end{equation} 
Notice that in our case,  the divergence is basically of order $\alpha/k^{s}$.

The 3--D Fourier transform in spherical coordinates can be written as \cite{MC,Grad}
\begin{equation}
V(r)=\alpha \Bigl(\frac{2}{\pi}\Bigr)^{1/2} \int_{0}^{p_{m}}  \frac{k \sin(kr) }{(\hbar k)^{s} r} \,dk.
\end{equation}
Thus, the Fourier transform upon $1/ k^{s}$ can be expressed generically as follows
\begin{eqnarray}
\label{FT}
V(r)=\frac{\sqrt{\frac{2}{\pi }} p_{m}^{3-s} \, _1F_2\left(\frac{3}{2}-\frac{s}{2};\frac{3}{2},\frac{5}{2}-\frac{s}{2};-\frac{1}{4} p_{m}^{2} r^2\right)}{3-s},
\end{eqnarray}
where $_{1}F_{2}$ is a generalized hypergeometric function valid for $0<s<3$. The term $p_{m}$ is a maximum momenta, i.e., a cutoff, included in order to avoid the divergence of $V(r)$, clearly small enough compared with $p_{0}$.
Fortunately, $_{1}F_{2}\rightarrow \frac{3 p_{m}^{8/3}}{4 \sqrt{2 \pi }}$ when $r \rightarrow 0$, i.e., is regular at the origin. Consequently, the pseudo potential is just a simple delta--function potential $ U_{P}(r)=U_{0}\delta(r)$ for $0<s<3$, see for instance Ref.\,\cite{maeda} 

It is noteworthy to mention that even in this situation the action of the operator (\ref{PP}), removes the divergence in the Fourier transform (\ref{FT}).
Thus, by subtracting the divergent term $(U_{0}N/V)^{2}/2\epsilon_{k}$ in the corresponding Hamiltonian (\ref{Ham3}), we safely remove the divergent behavior.

Finally, our Hamiltonian now is given by
\begin{eqnarray}
\label{Ham301}
\hat{H}& =& \frac{U_0N^2}{2V}+\sum_{\vec{k}\not=0}\sqrt{\epsilon_{k} \Bigl(\epsilon_{k}+\frac{2U_0N}{V}\Bigl)} 
\hat{b}_{\vec{k}}^{\dagger}\hat{b}_{\vec{k}}\,\,\,\,\,\,\,\,\,\,\,\,
\\\nonumber&+&\sum_{\vec{k}\not=0}\Bigg\{-\frac{1}{2}\Bigg[\frac{U_0N}{V} +\epsilon_{k}
-\sqrt{\epsilon_{k}\Bigl(\epsilon_{k}+\frac{2U_0N}{V}\Bigr)}\\\nonumber&-& \Bigl(\frac{U_0N}{ V}\Bigr)^2\frac{1}{2 \epsilon_{k}}\Bigg]\Bigg\}. 
\end{eqnarray}
Thus, in one hand, the ground state energy is well defined for values $0<s<3$ compatible with the result Eq.\,(\ref{eq:TC}).
On the other hand, from the $N$--body Hamiltonian (\ref{Ham301}), we are able to recognize the energy of the Bogoliubov excitations
\begin{equation}
\label{Ek}
E_{k} = \sqrt{\epsilon_{k}\Bigl(\epsilon_{k}+
 \frac{2U_0N}{V}\Bigr)}. 
\end{equation}
The long--wavelength limit, $k\rightarrow 0$, associated with the above expression, leads to the following  dispersion relation %
\begin{equation}
\label{qpart}
E_{p}=\frac{p^{s/2}}{\alpha^{1/2}}\sqrt{\frac{2U_{0}N}{V}}\left[1+\frac{p^{s}}{4 \alpha U_{0} N}\right]
\end{equation}

Notice that in the case $s=2$ we recover the linear phonon--like dispersion relation for the excitations, i.e., $E_{k} \sim p$. 

However, in the case $s \sim 1/3$ we obtain 
\begin{equation}
E_{p}=\frac{p^{1/6}}{\alpha^{1/2}}\sqrt{\frac{2U_{0}N}{V}}\left[1+\frac{p^{1/3}}{4 \alpha U_{0} N}\right]
\end{equation}
which basically behaves as $ E_{k} \sim p^{1/6}$, for low momenta valid up to some value $p_{m}$.


\end{document}